\documentclass{PoS}

\def\lsim{\raise0.3ex\hbox{$<$\kern-0.75em\raise-1.1ex\hbox{$\sim$}}}
\def\gsim{\raise0.3ex\hbox{$>$\kern-0.75em\raise-1.1ex\hbox{$\sim$}}}

\title{Continuum extrapolation of finite temperature meson correlation
functions in quenched lattice QCD}

\ShortTitle{Continuum extrapolation of finite temperature quenched lQCD}

\author{Anthony Francis\thanks{Poster contribution.}\\
        Fakult\"at f\"ur Physik, Universit\"at Bielefeld, D-33615 Bielefeld, Germany\\
        E-mail: \email{afrancis@physik.uni-bielefeld.de}}

\author{
        Frithjof Karsch\thanks{Speaker.}\\
        Physics Department, Brookhaven National Laboratory, Upton, NY 11973, USA\\
        E-mail: \email{karsch@bnl.gov}}

\abstract{We explore the continuum limit $a\rightarrow 0$  of meson 
correlation functions at finite temperature. In detail we analyze finite 
volume and lattice cut-off effects in view of possible consequences for 
continuum physics. We perform calculations on quenched gauge configurations 
using the clover improved Wilson fermion action. 
We present and discuss simulations on isotropic $N_\sigma^3\times 16$ lattices 
with $N_\sigma=32,48,64,128$ and $128^3 \times N_\tau$ lattices with 
$N_\tau=16,24,32,48$ corresponding to lattice spacings in the range of 
$0.01\ {\rm fm}\ \lsim\ a \ \lsim\ 0.031 \ {\rm fm}$ at $T\simeq1.45T_c$.
Continuum limit extrapolations of vector meson and pseudo scalar correlators 
are performed and their large distance expansion in terms of thermal moments 
is introduced. We discuss consequences of this analysis for the calculation of 
the electrical conductivity of the QGP at this temperature. 

          }

\FullConference{The XXVIII International Symposium on Lattice Field Theory, Lattice2010\\
		June 14-19, 2010\\
		Villasimius, Italy}

\begin{document}

\kopfauthoren{A. Francis, F. Karsch}

\section{Introduction}
The heavy-ion experiments at RHIC and soon LHC are probing deeper and deeper 
into the high temperature region of QCD. As such more and more
experimental results are presented that require a more complete knowledge of 
finite temperature QCD.
Naturally one turns to lattice QCD to give some answers to the questions 
arising. However some questions of interest are notoriously difficult
to answer from a lattice perspective. This is in particular the case 
for the analysis of spectral properties of hadron correlation functions.
Although here the main interest is in extracting information on hadronic
spectral functions at low frequencies, this cannot be achieved without
good control over the large frequency region. Correlation functions 
thus need to be controlled at short as well as large distances.
High accuracy data on large lattices at several values of the cut-off are 
needed to control finite volume and lattice cut-off
effects. Only then a reliable continuum extrapolation becomes possible 
which is a prerequisite for extracting dependable physics results. 

In this combined talk and poster contribution we present results from a 
lattice analysis of meson correlation functions at finite 
temperature \cite{karsch-francis}. We emphasize here our systematic
analysis of the finite volume and cut-off dependence of thermal 
meson correlation functions. This analysis 
leads to the conclusion that we indeed can take the continuum
limit for the vector and pseudo scalar current correlation functions in
a large Euclidean time interval. 
We also calculate several thermal moments of hadron spectral functions
to better explore the spectral properties at low frequencies. 
We close with a discussion of the constraints arising from our analysis
for the determination of the electrical conductivity in the quark gluon plasma 
at finite temperature \cite{gupta, aarts-kim}.

\section{Meson Correlation Functions}
Our key observable of interest is the Euclidean correlation function of a 
given particle current,
\begin{eqnarray}
 J_{\nu}\equiv\bar q(\tau,\vec x)\gamma_{\nu} q(\tau,\vec x)\,\,,
\label{eqn-current}
\end{eqnarray}
where choosing the appropriate gamma matrix we obtain the vector particle channels for $\gamma_\nu=\gamma_\mu$ where $\mu=0,...,4$ and 
the pseudo scalar for $\gamma_\nu=\gamma_5$. 

In this work we analyze the Euclidean temporal two-point correlation function for the above channels at fixed momentum,
\begin{eqnarray}
 G_{\mu\nu}(\tau, \vec p)=\sum_{\vec x} G_{\mu\nu}(\tau,\vec x) e^{i\vec p \cdot \vec x}  ,
\end{eqnarray}
 where 
\begin{eqnarray}
 G_{\mu\nu}(\tau,\vec x)=\langle J_{\mu}(\tau,\vec x)J^\dagger_{\nu}(0,\vec 0) \rangle ,
\end{eqnarray}
while we denote thermal expectation values with $\langle ... \rangle$. 
In this work we will set the momentum to zero, $\vec p =0$, and suppress
the second argument in the hadronic correlation functions. 

At high temperatures the spectrum of meson resonances is more and more changed 
by thermal effects; the width of resonances will broaden and the onset of the 
continuum in the spectrum may shift. As
a result the interest in the analysis of the above current-current correlation 
function shifts to their representation in terms of a spectral function 
$\rho_{\mu\nu}$ \cite{karsch-wyld}:
\begin{eqnarray}
 G_{\mu\nu}(\tau,T)=\int^\infty_0\,\frac{d\omega}{2\pi}\,\rho_{\mu\nu}(\omega,T)\,\frac{\cosh(\omega(\tau-1/2T))}{\sinh(\omega/2T)}.
\end{eqnarray}
From here on we denote with $\mu\nu\equiv ii$ the sum over the space-like 
components of the vector spectral function $\rho_{ii}$, while the full vector 
spectral function is given by $\rho_V\equiv\rho_{00} + \rho_{ii}$. We also
use the notation $PS$ for $\mu\nu\equiv 55$.
Note that the correlation function $G_{00}(\tau,T)$ is connected to the net 
number of quarks $(q-\bar q)$ in a given flavor channel, {\it i.e.}, 
$n_0(\tau)=\int\,d^3x\,J_0(\tau,\vec x)$. 
As the net quark number is conserved, it does not depend on time, 
$n_0(\tau)=n_0$. Thus the corresponding 
correlation function is constant in Euclidean time and the spectral function is simply given by,
\begin{eqnarray}
 \rho_{00}(\omega)=-2\pi\chi_q\omega\delta(\omega)\,\,,
\end{eqnarray}
with the quark number susceptibility
\begin{eqnarray}
 \chi_q=-\int\,d^3x\,\langle J_0(\tau,\vec x)J^\dagger_0(0,\vec 0)\rangle\;.
\end{eqnarray}
Consequently the time-like correlation function obeys the relation 
$G_{00}(\tau,T)\equiv -\chi_qT$, which immediately leads to an exact relation 
between $G_{ii}(\tau,T)$ and $G_{V}(\tau,T)$:
\begin{eqnarray}
 G_{ii}(\tau,T)=\chi_qT+G_{V}(\tau,T)\,\, .
\end{eqnarray}

In the free field limit the spectral functions are seen to increase 
quadratically for large frequencies $\omega$. One obtains:
\begin{eqnarray}
 \rho_{ii}(\omega)&=&2\pi \chi_q\omega\delta(\omega) + \frac{3}{2\pi}\omega^2\tanh(\omega/4T)\\
 \rho_{V}(\omega)&=&\rho_{00}(\omega)+\rho_{ii}(\omega)=\frac{3}{2\pi}\omega^2\tanh(\omega/4T)\\
 \rho_{PS}(\omega)&=&\frac{3}{4\pi}\omega^2\tanh(\omega/4T)
\end{eqnarray}
At finite temperature the exact cancellation of the $\delta$-functions in the $\rho_V(\omega)$ spectral function is only approximate, as 
$\rho_{00}(\omega)$ continues to be proportional to a $\delta$-function  due to 
the connection with net quark number, while the $\delta$-function in 
$\rho_{ii}(\omega)$ is subject to thermal effects. As a consequence this 
contribution is smeared out into a Breit-Wigner shaped peak \cite{aarts-resco}:
\begin{eqnarray}
 \rho_{ii}(\omega)\rightarrow\rho_{ii}^{BW}(\omega)=2\chi_q\; c_{BW}\frac{\omega\Gamma/2}{\omega^2+(\Gamma/2)^2}+(1+\kappa)\cdot\frac{3}{2\pi}\omega^2\tanh(\omega/4T)\,\,,
\label{Ansatz}
\end{eqnarray}
where $\kappa$ parametrizes corrections to the high frequency behavior of 
the free field limit.
This smearing in the low frequency region is directly related to the 
appearance of finite transport coefficients in the thermal medium, 
e.g. the electrical conductivity,
\begin{eqnarray}
 \frac{\sigma}{T}=\frac{C_{em}}{6}\lim_{\omega\rightarrow0}\frac{\rho_{ii}(\omega)}{\omega T}\,\, ,
\label{eq-electrical}
\end{eqnarray}
where $C_{em}$ is the sum over the squared charges of the contributing  quark
flavors. The pseudo scalar spectral function on the other hand does not per se 
contain an additional low frequency contribution; a connection with 
transport phenomena is not expected.

\section{Thermal Moments of the Correlation Function}
A useful set of observables that helps to characterize the structure
of spectral functions, is given  by ``thermal moments'' of the spectral 
function at vanishing momentum. At a given order we define these moments as 
the Taylor expansion coefficients of the correlation function expanded 
around the midpoint of the Euclidean time interval\footnote{As the spectral
functions as well as the integration kernel in the spectral representation
of the correlators are odd functions of the frequency, all odd moments vanish.},
{\it i.e.}, 
around $\tau T$ = 1/2,
\begin{eqnarray}
G_H^{(2n)}=\left.\frac{1}{(2n)!}\frac{d G_H(\tau T)}{d (\tau T)^{2n}}\right|_{\tau T=1/2}=\frac{1}{(2n)!}\int_0^\infty\,\frac{d\omega}{2\pi}\left(\frac{\omega}{T}\right)^{2n}\,\frac{\rho_H(\omega)}{\sinh(\omega/2T)}
\label{Taylor}
\end{eqnarray}
and
\begin{eqnarray}
G_H(\tau T)=\sum_{n=0}^\infty \,G_H^{(2n)}\,\left( \frac{1}{2}-\tau T \right)^{2n}.
\end{eqnarray}

The infinite temperature, free field limit of the correlation function can be calculated analytically. For massless quarks one obtains \cite{florkowski}:
\begin{eqnarray}
 G_V^{free}(\tau T) =2 G_{PS}^{free}(\tau T) &=& 6T^3\left( \pi(1-2\tau T)\,\frac{1+\cos^2(2\pi\tau T)}{\sin^3(2\pi\tau T)}+2\,\frac{\cos(2\pi \tau T)}{\sin^2(2\pi\tau T)} \right)
\label{freeG}
\\
G_{ii}^{free}(\tau T) &=& T^3+G_V^{free}(\tau T) 
\end{eqnarray}
Using these results the first three non-vanishing moments are then given by 
\begin{eqnarray}
 G_V^{(0),free}&=& 2 G_{PS}^{(0),free} = \frac{2}{3}G_{ii}^{(0),free}=2T^3\;,\\
 G_H^{(2),free}&=& 2 G_{PS}^{(2),free} = \frac{28\pi^2}{5}T^3\;, \\
 G_H^{(4),free}&=& = 2 G_{PS}^{(4),free} = \frac{124\pi^4}{21}T^3 \; \;.
\end{eqnarray}
In our lattice simulation we will also analyze ratios of moments. In the
free field limit they are given by 
\begin{eqnarray}
R_{V,free}^{(2,0)}&=&  \frac{G_V^{(2),free}}{G_V^{(0),free}} = R_{PS,free}^{(2,0)}= \frac{14\pi^2}{5}\simeq27.635\\
R_{ii,free}^{(2,0)}&=&\frac{2}{3}R_{V,free}^{(2,0)}\simeq18.423\\
R_{H,free}^{(4,2)}&=&\frac{155\pi^2}{147}\simeq10.407\;\;,
\end{eqnarray}
which can be obtained from the ratio of the mid-point subtracted correlation 
functions and the corresponding free field values:
\begin{eqnarray}
\frac{\Delta_H(\tau T)}{\Delta_H^{free}(\tau T)}= \frac{G_H(\tau T)-G_H^{(0)}}{G_H^{free}(\tau T)-G_H^{(0),free}}
=\frac{G_H^{(2)}}{G_H^{(2),free}}\left( 1+(R_H^{(4,2)}-R_{H,free}^{(4,2)})(\frac{1}{2}-\tau T)^2+ ... \right)\,.
\label{moment}
\end{eqnarray}
Note that the distinction between $H=ii$ or $H=V$ is unnecessary when 
evaluating $\Delta_H(\tau T)$, as differences in hte correlators, 
which arise from 
contributions of $\delta$-functions in the spectral functions, are eliminated 
in subtracted correlators.

\section{Simulation Parameters and Results}

\subsection{Computational Details}
Our numerical results are obtained from quenched QCD gauge field configurations generated with the standard SU(3) single plaquette Wilson gauge action \cite{wilson}. Using an 
over-relaxed heat bath algorithm configurations were generated on lattices of size $N_\sigma^3 \times N_\tau$, where
$32 < N_\sigma < 128$ and $N_\tau = 16$, 24, 32 and 48, with a separation of 500 updates per configuration. Correlation functions and plaquette expectation values calculated on these configurations have been checked and are 
seen to be uncorrelated. Details on the statistics collected are given in Tab.~\ref{all-par}.

All calculations presented here have been performed at a single temperature
value, $T\simeq 1.45T_c$. The gauge couplings have  been adjusted accordingly
for the different temporal lattice sizes. To do so
we extrapolate known results for the critical coupling $\beta_c(N_\tau)$ and 
the square root of the string tension $\sqrt{\sigma}$ \cite{lucini}. 
This is achieved by fitting $T/\sqrt{\sigma}$ using the ansatz suggested 
in \cite{allton} within the range of $\beta\in[5.6,6.5]$ and extrapolating to 
our relevant region of $\beta\in[6.8,7.8]$. All simulation parameters are 
given in Tab.~\ref{all-par}.

In the fermion sector we employ the clover improved Wilson action with non-perturbatively chosen clover coefficient $c_{SW}$ \cite{luescher1} and determine the correlation functions using an even-odd decomposed
CG-algorithm. The hopping parameter $\kappa$ is chosen to be close to its critical value \cite{luescher2} and tuned to give approximately constant quark masses for four of the examined 
lattices. Here the quark masses are estimated using the axial Ward identity (AWI) to calculate the AWI current quark mass $m_{AWI}$ 
for the different values of cut-off.
Whereby we use the non-perturbatively improved axial vector current with coefficient $c_A$ noted in \cite{sommer}. In the next step we calculate the renormalization
group invariant mass $m_{RGI}$ and rescale to the commonly quoted 
$\overline{MS}$-scheme at the scale $\mu=2$GeV. The results are also listed 
in Tab.~\ref{all-par}.

\begin{table}
\begin{center}
\begin{tabular}{|c|c|c|c|c|c|c|c|c|c|c|}
\hline
$N_\tau$ & $N_\sigma$ & $\#$ conf &$\beta$ & $a[\textrm{fm}]$ &$T/T_c$ & $c_{SW}$ & $\kappa$ & $m_{\overline{MS}}/T_{[\mu=2\textrm{{\tiny GeV}}]}$ &$Z_V$ & $Z_{PS}$ \\\hline
48& 128 & 451 &7.793 & 0.010 &1.43 & 1.3104 & 0.13340 & 0.1117(2)&0.861 & 0.79 \\\hline
32& 128 & 255 &7.457 & 0.015 &1.45 & 1.3389 & 0.13390 & 0.0989(4)&0.851 & 0.78 \\\hline
24& 128 & 340 &7.192 & 0.021 &1.42 & 1.3673 & 0.13431 & 0.1062(2)&0.842 & 0.76 \\
& & 156& & & & & 0.13440&0.02367(5)&&\\\hline
16& 128 & 191 &6.872 & 0.031 &1.46 & 1.4125 & 0.13495 & 0.02429(5)&0.829 & 0.74 \\
& 64 & 191 & & & & & & & & \\
& 48 & 229 & & & & & & & & \\
& 32 & 251 & & & & & & & & \\\hline
\end{tabular}
\end{center}
\caption{Simulation parameters for the generation of gauge field configurations on
lattices of size $N_\sigma^3\times N_\tau$.}
\label{all-par}
\end{table}



\subsection{Numerical Results}

\begin{figure}
\includegraphics[width=.5\textwidth]{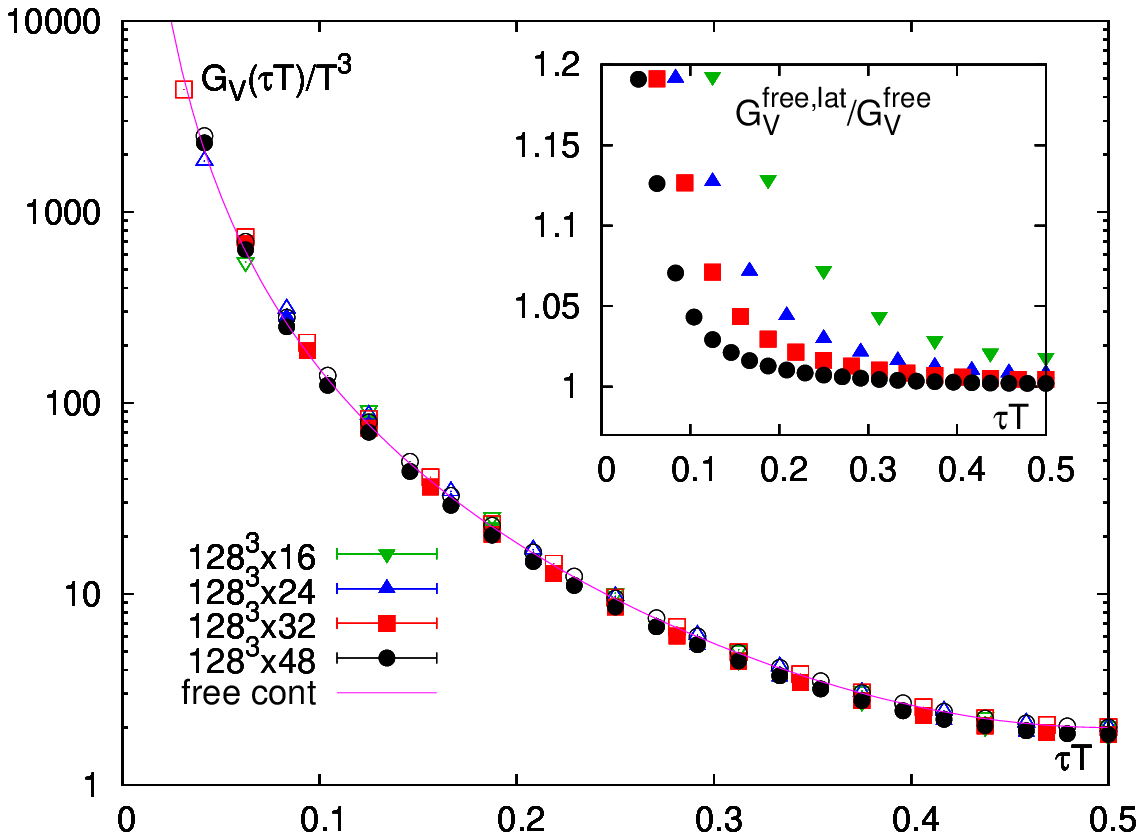}
\includegraphics[width=.5\textwidth]{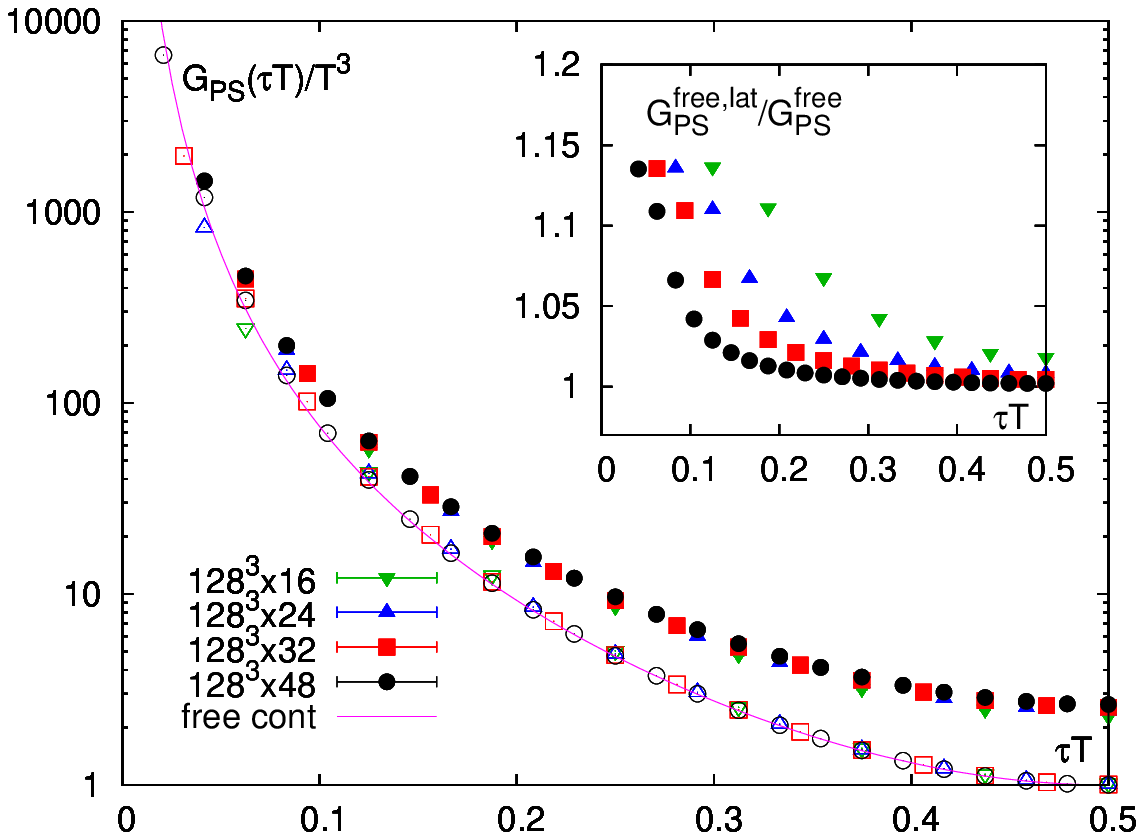}
\caption{The vector (left) and pseudo scalar (right) correlation functions together with their free continuum (magenta lines) and free lattice 
(open symbols) counterparts. (Blow-ups): The ratio of free lattice and free continuum correlation functions.}
\label{freecorrs}
\end{figure}

Due to the subtlety of the effects in question it is not very illuminating to examine the correlator directly. To illustrate this
we show the correlation functions of the vector and pseudo scalar currents together with their analytically obtained free continuum
and free discretized, lattice counterparts in Fig.~\ref{freecorrs}. Clearly the exponential decay of the correlation function
dominates and obscures all other interesting physics features. 
To circumvent this effect it is useful to look at certain ratios that 
largely cancel the exponential part of the Euclidean time dependence of the
correlators. 


One of these ratios is that of the correlation function itself divided by 
the free correlation function given in Eq.~\ref{freeG}, 
\begin{eqnarray}
 R_H(\tau, T)=\frac{G_H(\tau, T)}{G^{free}_H(\tau, T)} \,\,.
\label{eq-ratio1}
\end{eqnarray}
The immediate advantage of a ratio as that given in Eq.~\ref{eq-ratio1} 
is that it gives a direct handle on the deviation of the correlation function 
calculated at finite and infinite temperature, respectively. As the 
continuum as well as lattice free correlation functions are known 
analytically, this also provides insight into the importance
of lattice cut-off effects as function of Euclidean time. 
The insertions in Fig.~\ref{freecorrs} show the ratio of the free lattice   
and the free continuum correlation functions. They, of course, show that 
cut-off effects are largest at small distances.
Moreover, it becomes obvious that with increasing cut-off values 
(larger $N_\tau$) the onset of cut-off effects is shifted to 
smaller distances.

As we employ the local current correlations introduced in Eq.~\ref{eqn-current} all results have to be renormalized multiplicatively, 
\begin{eqnarray}
J_{\nu}^{lat}=(2\kappa Z_H) \bar q(\tau,\vec x)\gamma_{\nu} q(\tau,\vec x), 
\end{eqnarray}
where the renormalization constants have been determined non-perturbatively for the vector \cite{luescher2} and perturbatively up to two-loop order for the 
pseudo scalar channel \cite{greeks}. They are listed in Tab.~\ref{all-par}.\\
In the case of the vector correlation functions the above statement holds 
true also for the time-like component $G_{00}(\tau,T)$ because the local 
lattice current is not conserved at non-zero 
lattice spacing. Consequently the time-like correlation function  
$G_{00}(\tau, T)$ may not be strictly $\tau$-independent. 
However, as all vector current-current correlation functions are subject to 
the same renormalization 
constants a rescaling of the correlation functions $G_{V}(\tau,T)$ and 
$G_{ii}(\tau,T)$ or 
even the ratio $R_H(\tau,T)$ by $G_{00}(\tau,T)$ yields a quantity independent 
of renormalization.

\subsection{Finite Volume and Lattice Effects}

\subsubsection{Quark Number Susceptibility}
\begin{figure}
\includegraphics[width=.5\textwidth]{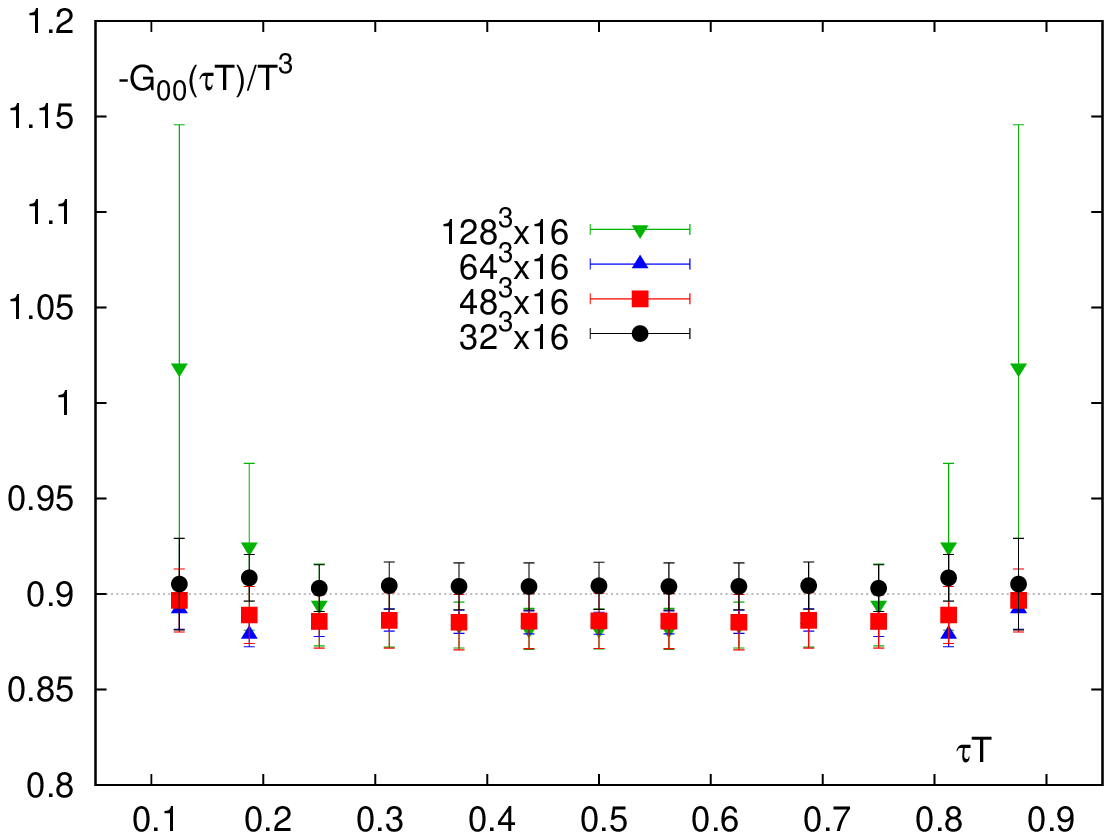}
\includegraphics[width=.5\textwidth]{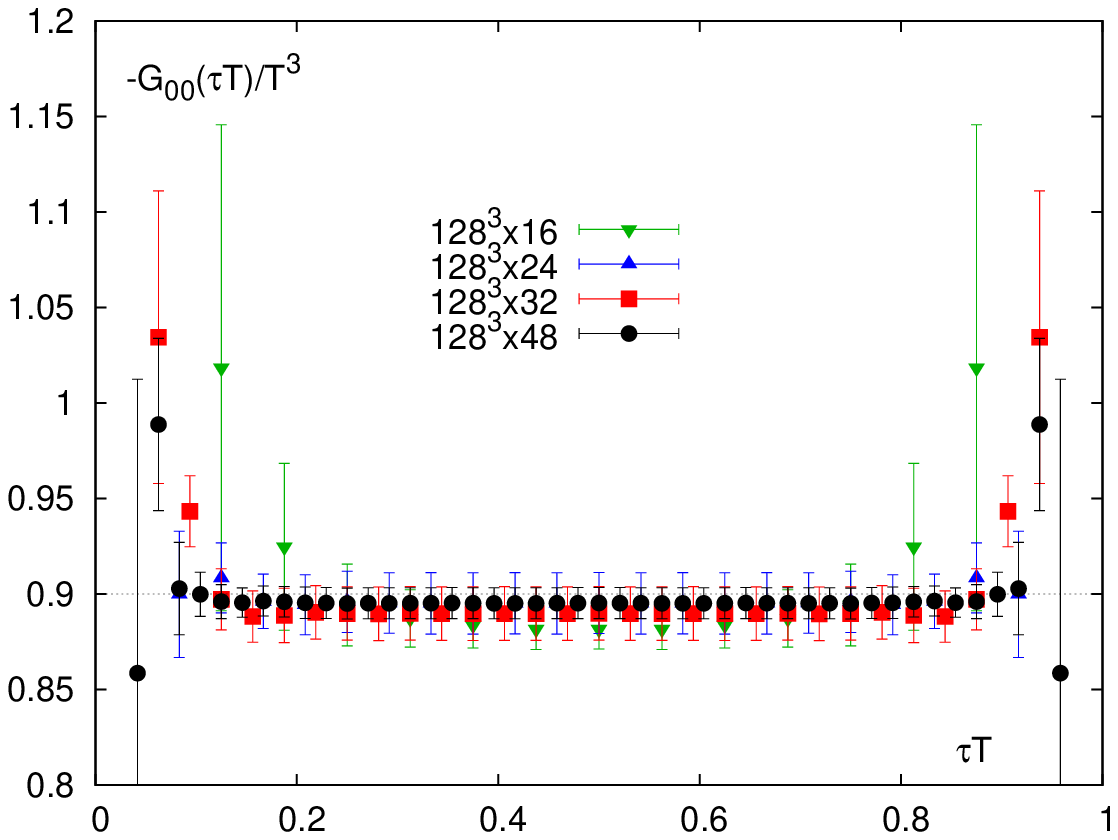}
\caption{The time-like component of the vector spectral function, $G_{00}(\tau T)/T^3$,
calculated at $T\simeq1.45T_c$. The left hand part of the figure
shows the volume dependence of $G_{00}(\tau T)/T^3$ for $N_\tau = 16$ and $32 \le N_\sigma \le 128$. The
right hand figure shows the cut-off dependence of $G_{00}(\tau T)/T^3$ for $N_\sigma = 128$ and
$16 \le N_\tau \le 48$.}
\label{v4channel}
\end{figure}

As mentioned above the time-like component of the vector correlation function,
$G_{00}(\tau T)$, will be used for rescaling in the following, so it is the first channel to be tested for finite volume 
and cut-off effects. We find these effects to be small, as can be seen from 
Fig.~\ref{v4channel}, where we plot $-G_{00} (\tau T)/T^3$. The
quark number susceptibility may then be extracted from the long distance 
behavior of the correlator, $\chi_q /T^2 = -G_{00}(\tau T \simeq 0.5)/T^3 $.

On the left of Fig.~\ref{v4channel} we show the results for the $N_\tau=16$ lattice with fixed cut-off while varying the spatial extent. Except for
aspect ratio $N_\sigma/N_\tau=2$ all results agree within statistical errors 
at the level of $1\%$. On the right the cut-off dependence at fixed spatial 
extent $N_\sigma=128$ is shown. Here too systematic effects in the quark 
number susceptibility are seen to be small.

Note even though we implement the non-conserved local current of the time-like vector correlation function, we obtain results that are $\tau$-independent to a
very high degree. The deviations at small distances $\tau T$ might be 
understood as lattice effects. 
The results for the quark number susceptibility are summarized in Tab.~\ref{v4ch-tab}.

\begin{table}
\begin{center}
\begin{tabular}{|c|c|c|c|c|c|}
\hline
$N_\tau$&16&24& 32 &48 &$\infty$ \\ \hline
$\chi_q/T^2$& 0.882(10)&0.895(16)&0.890(14)&0.895(8)&0.897(3)    \\\hline
\end{tabular}
\end{center}
\caption{Quark number susceptibility ($\chi_q/T^2$) calculated on lattices of size $128^3\times N_\tau$ .
The quark number susceptibilities have been renormalized using the renormalization
constants listed in Tab.~1. The last column gives the result of a continuum
extrapolation taking into account cut-off errors of $\mathcal{O}(a^2)$.}
\label{v4ch-tab}
\end{table}

\subsubsection{Vector and Pseudo Scalar Correlation Functions}
Turning to the vector and pseudo scalar correlation functions we stress that 
the large range of available spatial lattice sizes at fixed cut-off, with 
aspect ratios ranging between $2\le N_\sigma/N_\tau \le 8$, allows to 
quantify finite volume effects. Moreover, the large
temporal extent of maximum $N_\tau=48$ on an isotropic (!) lattice reduces
the lattice spacing to $0.01$fm at $T \simeq 1.45T_c$. The
variation of $N_\tau$ by up to a factor of three allows 
to control lattice cut-off effects. 
In the following it will be shown that finite volume effects in the 
correlation functions are well under control and a controlled extrapolation
to the continuum limit is indeed possible in a large Euclidean time interval. 
Note that that our current analysis 
improves on systematic errors that were present in earlier calculations 
of the vector spectral function
performed by employing the same discretization scheme \cite{karsch-wetzorke}.

\begin{figure}
\includegraphics[width=.5\textwidth]{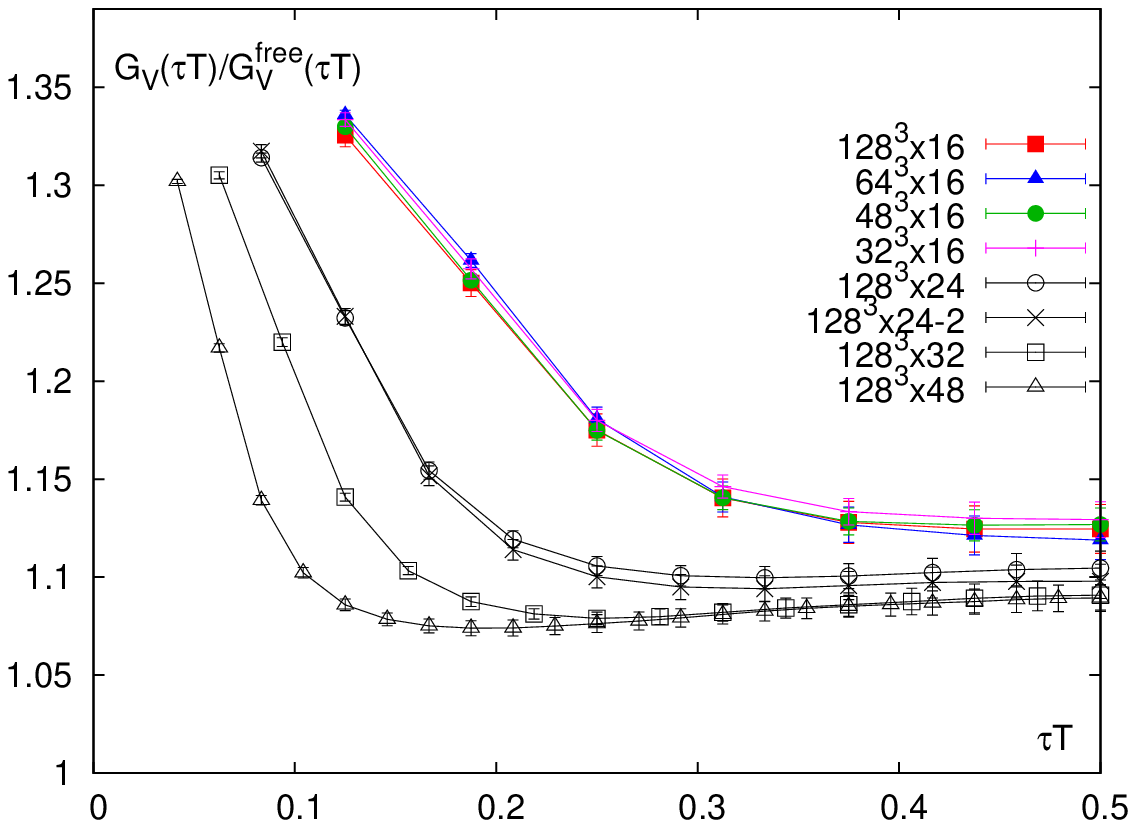}
\includegraphics[width=.5\textwidth]{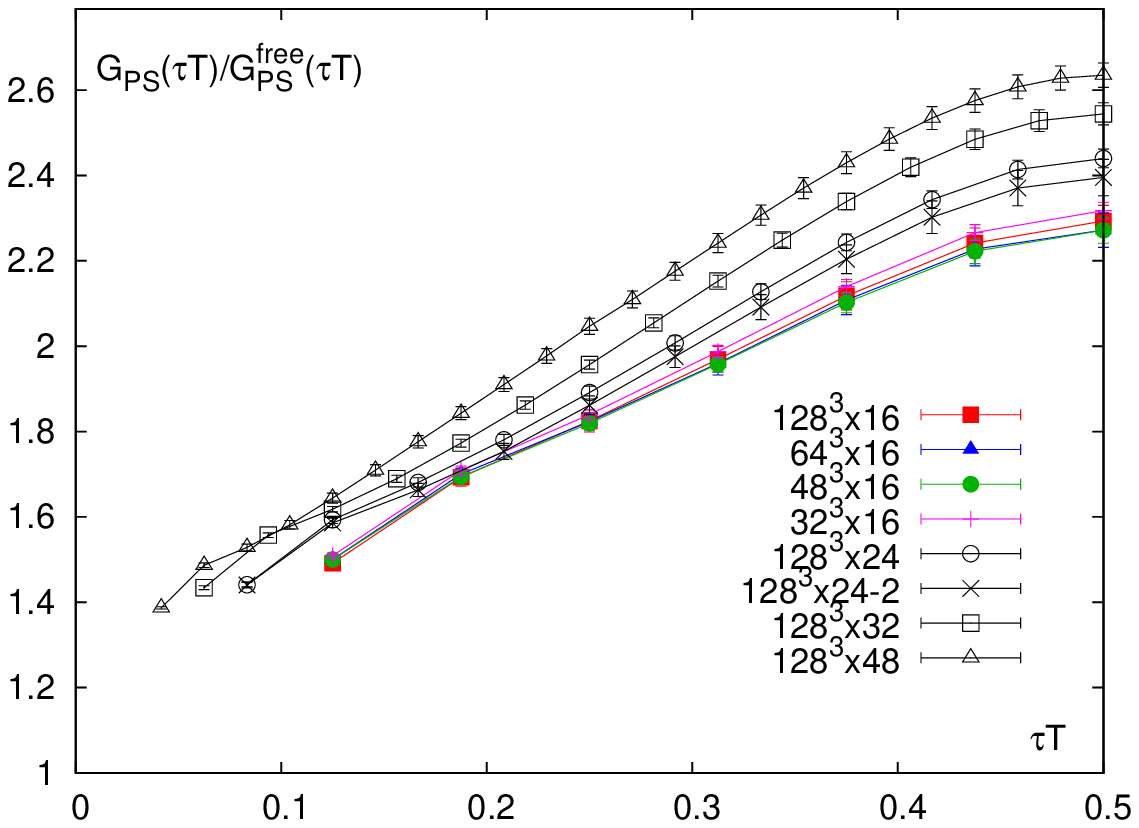}
\caption{The vector (left) and pseudo scalar (right) correlation functions, 
calculated on lattices of size $N_\sigma^3 \times N_\tau$
at $T\simeq1.45T_c$, where ``$128^3\times 24-2$`` denotes the lower quark mass on this lattice. The results are normalized using the corresponding free correlation functions. 
Shown are data for $\tau T > 1/N_\tau$ only.}

\label{volume-effects}
\end{figure}

In Fig.~\ref{volume-effects} we show results for the vector and pseudo scalar
correlation functions 
using the ratios introduced in Eq.~\ref{eq-ratio1}.
On the left hand side the ratio for $H=V$ and on the right hand side for $H=PS$
is shown for all available lattice sizes.
Data sets with fixed spatial size at $N_\sigma=128$ while varying 
the cut-off $N_\tau$ shown in black. Data sets with fixed cut-off 
($N_\tau=16$) and varying volume are shown in color. For one value of the 
cut-off  ($N_\tau=24$) we performed calculations for two different values of 
the quark masses.
We find  that finite quark mass effects are small and well within $2\%$.
From the fixed cut-off (colored) $N_\tau=16$ results in both plots finite 
volume effects for $\tau T \ge 0.3$ are seen to remain within a few percent 
even for the largest Euclidean time separation at $\tau T=0.5$. As a 
consequence these results show that finite volume effects are under control.
For $\tau T < 0.3$ in the left $H=V$ plot it is immediately apparent that 
cut-off effects are large in the ratio. 

\begin{figure}
\includegraphics[width=.5\textwidth]{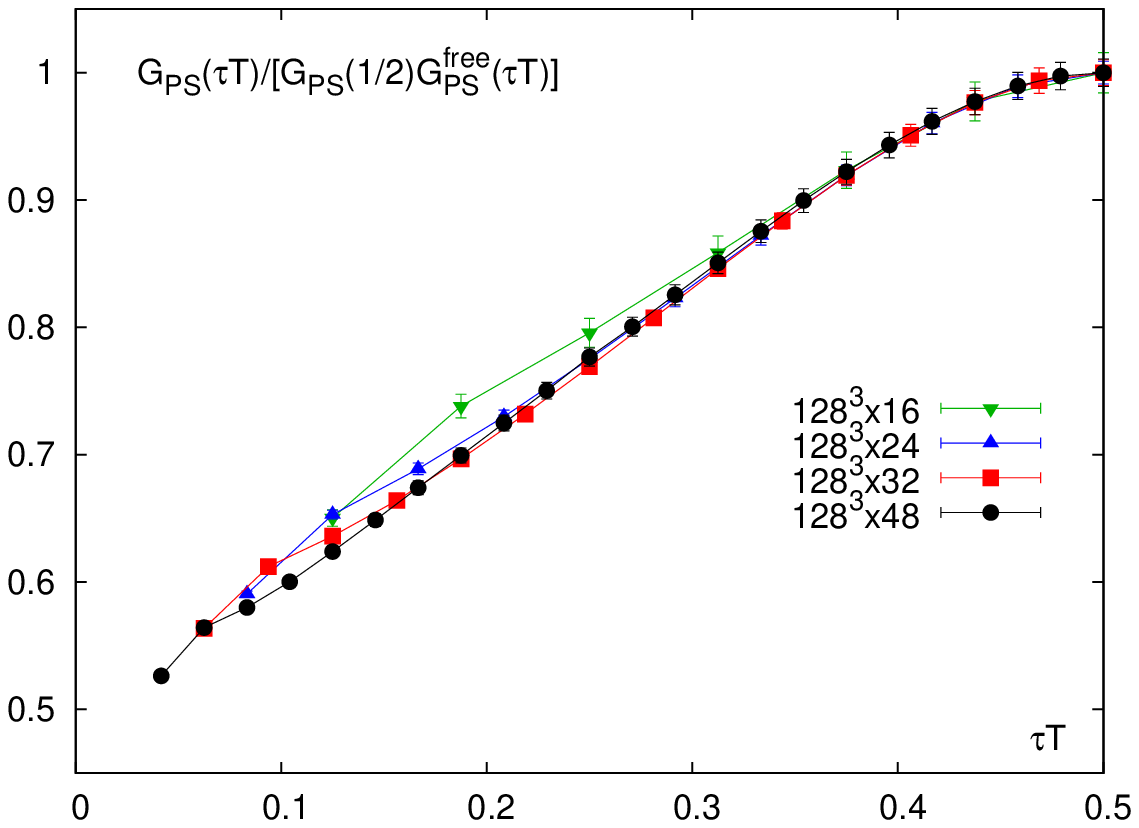}
\includegraphics[width=.5\textwidth]{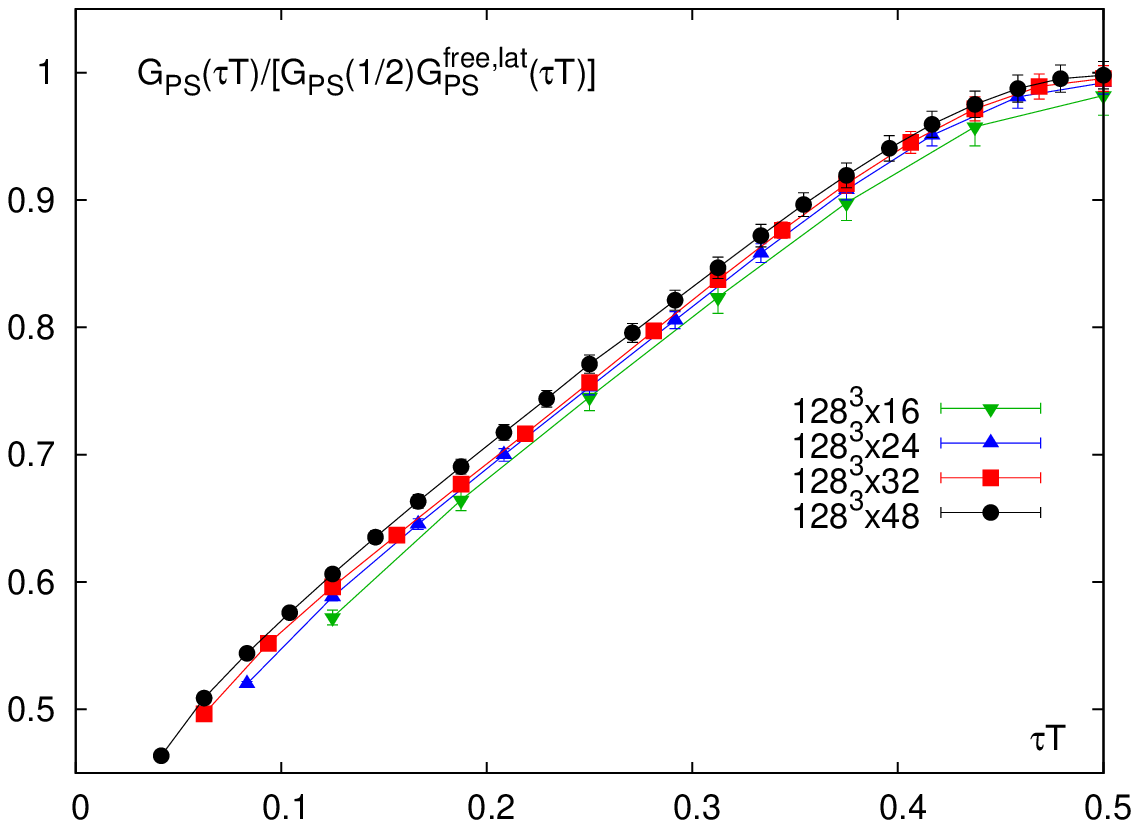}

\caption{The pseudo scalar correlation function normalized by the free continuum (left) and free lattice (right) correlation functions, both additionally rescaled 
by the correlator at $\tau T=1/2$, $G_{PS}(\tau T=0.5)$.}
\label{ps-effects}
\end{figure}

In the pseudo scalar case the situation concerning cut-off effects is not 
immediately evident. The ratio shown in Fig.~\ref{volume-effects}(right) 
shows large deviations from the free field behavior even at short distances.
At all distances the correlator thus seems to be controlled by large 
non-perturbative effects. 
Moreover, the analysis of cut-off effects is obscured by the fact that
data have been rescaled by using renormalization constants, which are
known only perturbatively. This introduces unknown uncertainties.
To eliminate at least these uncertainties 
we show in Fig.~\ref{ps-effects} the pseudo scalar 
correlation function normalized by the pseudo scalar correlation function
at $\tau T=0.5$. As we focus on the cut-off dependence we only show 
equal quark mass $N_\sigma=128$ results. The left hand figure 
shows the pseudo scalar correlator normalized by free continuum correlator
and in the right hand figure the free lattice correlation
function has been used. From the left hand plot the cut-off dependence becomes 
immediately evident and we conclude that, similarly to the vector case, cut-off
effects above $\tau T=0.3$ are small and increase with decreasing $\tau T$. 
Even though their effect becomes apparent for $\tau T\le0.3$, they do not 
dominate the behavior of the correlation function as in the vector channel. 
Actually the 
right hand side of Fig.~\ref{ps-effects} indicates that the $\tau$-dependence 
of the cut-off effects is similar to that of the free lattice correlation 
functions, as any cut-off effect is hardly visible also for  $\tau T \le 0.3$.

\subsection{Continuum Extrapolation} 
The results shown in the previous section indicate that the continuum limit
can be taken for the vector correlation functions at least for 
$\tau T \gsim 0.25$.
To do so we use a quadratic ansatz in $aT=1/N_\tau$ to fit 
ratios of the vector correlator and the corresponding free field values.
We normalize these ratios using the quark number susceptibility $\chi_q/T^2$
and perform fits at fixed temporal extent $\tau T$. As cut-off effects are
large on the $N_\tau=16$ lattice we will only use  data from the 
$N_\tau=24,\ 32,\ 48$ lattices, for reference we will however include the 
former in our figures.

\begin{figure}
\begin{center}
\includegraphics[width=.5\textwidth]{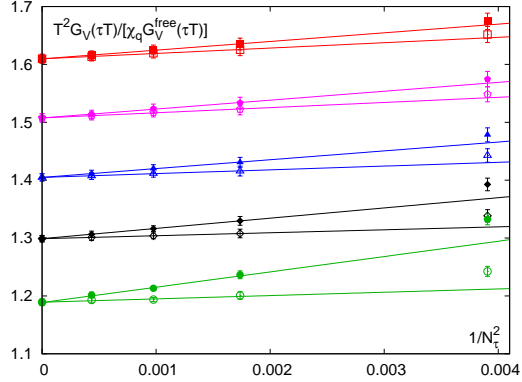}
\caption{The ratio $R_{V}(\tau T)$ normalized by the quark number susceptibility using the free continuum (full symbols) and lattice (open symbols) 
correlation functions over $1/N_\tau^2$. Shown are the results of $\tau T=1/2,7/16,3/8,5/16$ and $1/4$, for legibility the results have been offset by
$0.1$ respectively. For $\tau T=7/16$ and $5/16$ spline interpolations were used to estimate the corresponding results on the $N_\tau=24$ lattice.
}
\label{cont-extrapol}
\end{center}
\end{figure}

In Fig.~\ref{cont-extrapol} we show results of this extrapolation in 
$1/N_\tau^2$ for 
$\tau T=1/2,\ 7/16,\ 3/8,\ 5/16$ and $1/4$ where we used the free
continuum as well as the free lattice correlation functions for normalization. 
Note that we performed extrapolations for all Euclidean times
$\tau T$ available on the $N_\tau=48$ lattice. Wherever the smaller lattices 
fail to have a corresponding point in $\tau T$ we interpolate
using a spline construction. Subsequently the errors are then calculated 
using a Jackknife-method. In Fig.~\ref{cont-extrapol} this is the case
for $\tau T=7/16$ and $5/16$. The figure reveals that the continuum limit 
can be cleanly taken and consistent results are obtained by using the
free continuum and the free lattice normalizations, respectively. 

\begin{figure}
\includegraphics[width=.5\textwidth]{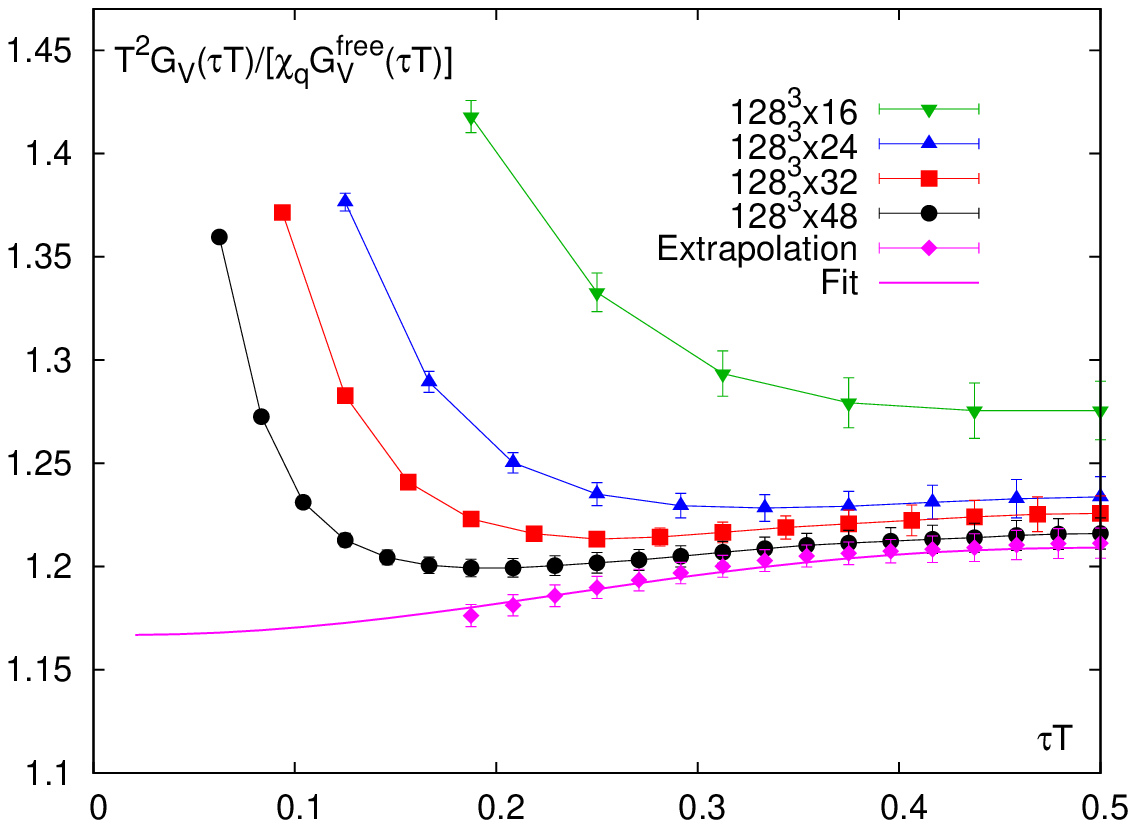}
\includegraphics[width=.5\textwidth]{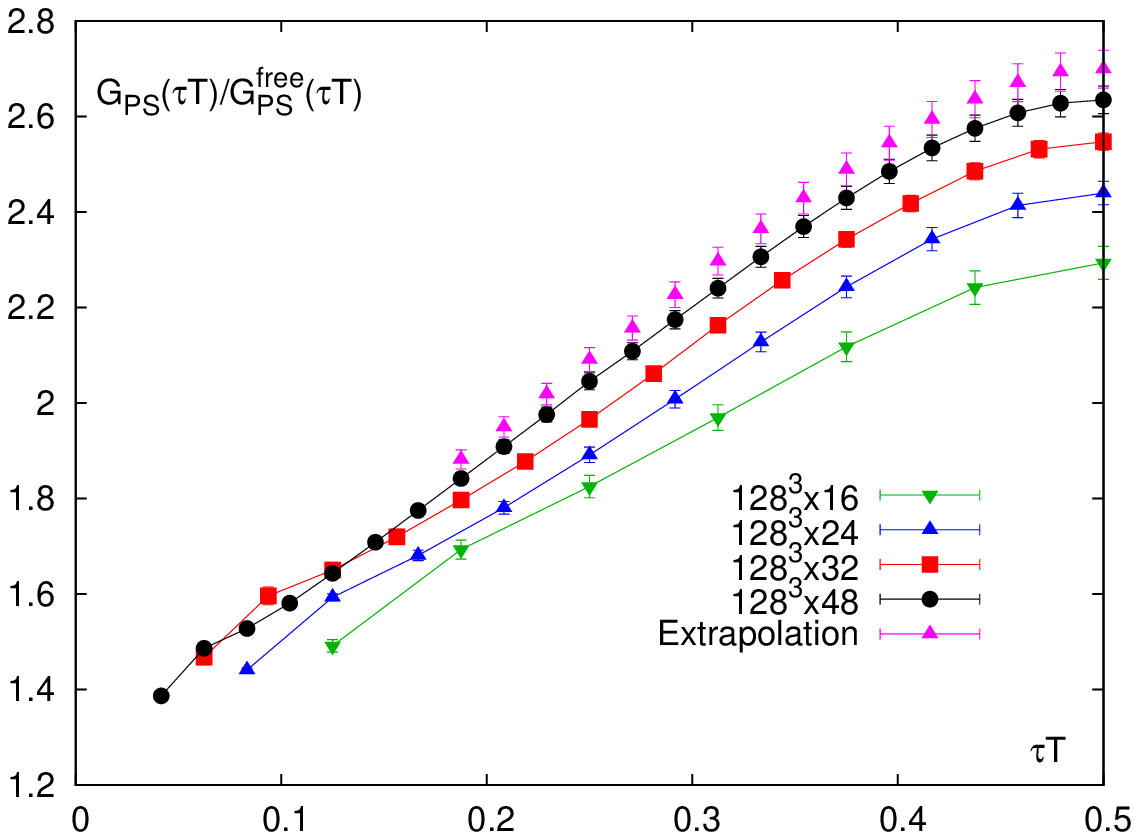}
\caption{(left): The ratio $R_{V}(\tau T)$ normalized by the quark number susceptibility using the free continuum correlation functions over $\tau T$ and 
its corresponding continuum extrapolation.
(right): The unnormalized $R_{PS}(\tau T)$ including its continuum extrapolation. In both cases the extrapolation was done as described, filling in 
spline interpolations when necessary. The solid curve in the left hand figure
shows the fit discussed in section 5.
}
\label{cont-corr}
\end{figure}

In the left hand part of Fig.~\ref{cont-corr} we show the results of the 
extrapolation in the vector channel as described above. 
Note that the largest deviation from the free correlation function occurs 
at $\tau T=1/2$. In fact, the established bending of the vector correlation 
function is crucial for a
quantitative description of the low frequency region of the vector spectral 
function in terms of the ansatz suggested in Eq.~\ref{Ansatz}. 
The short distance part of the correlation function obviously can be 
well described by the free spectral function including the correction factor 
$(1+\kappa)$ as also has been done in Eq.~\ref{Ansatz}.

In the right hand part of Fig.~\ref{cont-corr} the corresponding result for 
the pseudo scalar correlation function is shown. Here it is not possible to
suppress the renormalization effects using suitable  ratios of correlation
functions. The extrapolation necessarily also includes this ambiguity. 
As the correlator normalized by its value at the midpoint was
found to be almost cut-off independent and as finite volume effects were seen 
to be small renormalization effects dominate the uncertainty
of the extrapolation. 

\subsection{Curvature of the Vector and Pseudo Scalar Correlation Functions}
As discussed in section 3, thermal moments give additional insight into the
spectral representation of hadronic correlation functions. They are especially 
interesting as they are obtained at the largest Euclidean time separation
where the correlation functions are most sensitive to the low frequency region 
of the spectral function. In particular, the lower orders of the thermal
moments restrict the magnitude of the low frequency contribution to the 
spectral function and thus to the correlation function.

\begin{figure}
\includegraphics[width=.5\textwidth]{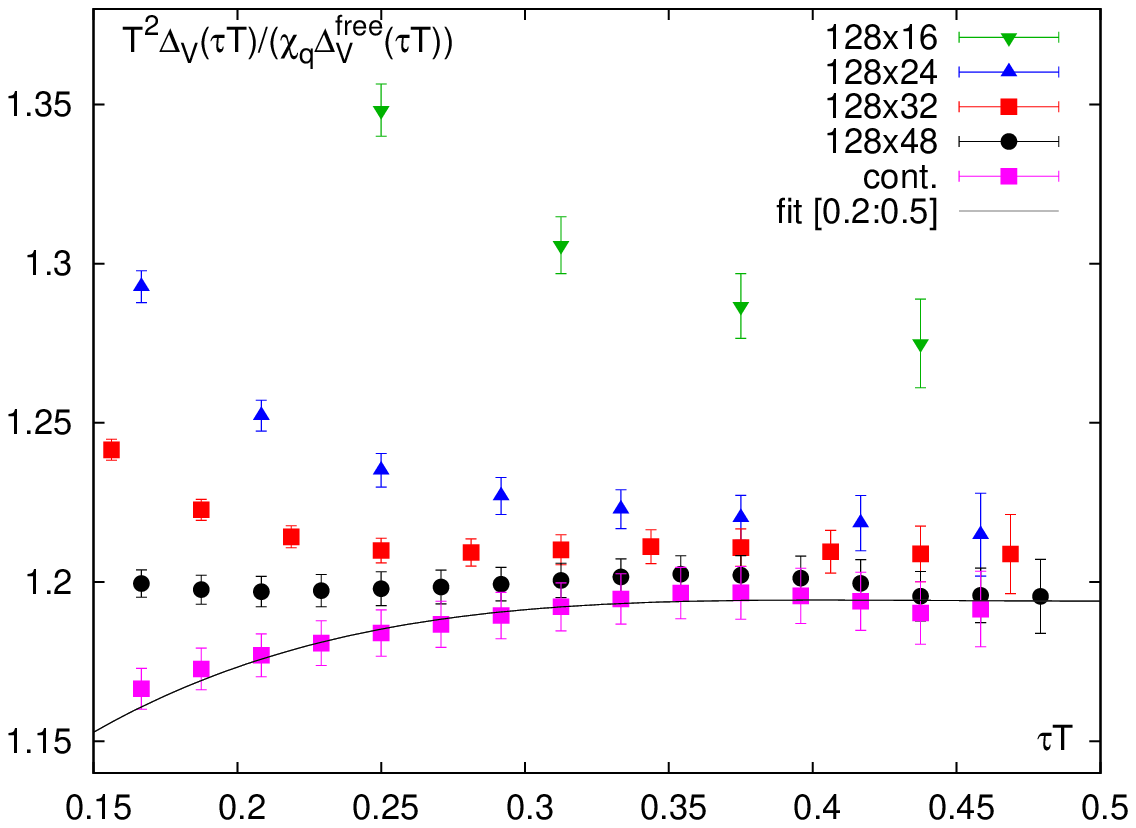}
\includegraphics[width=.5\textwidth]{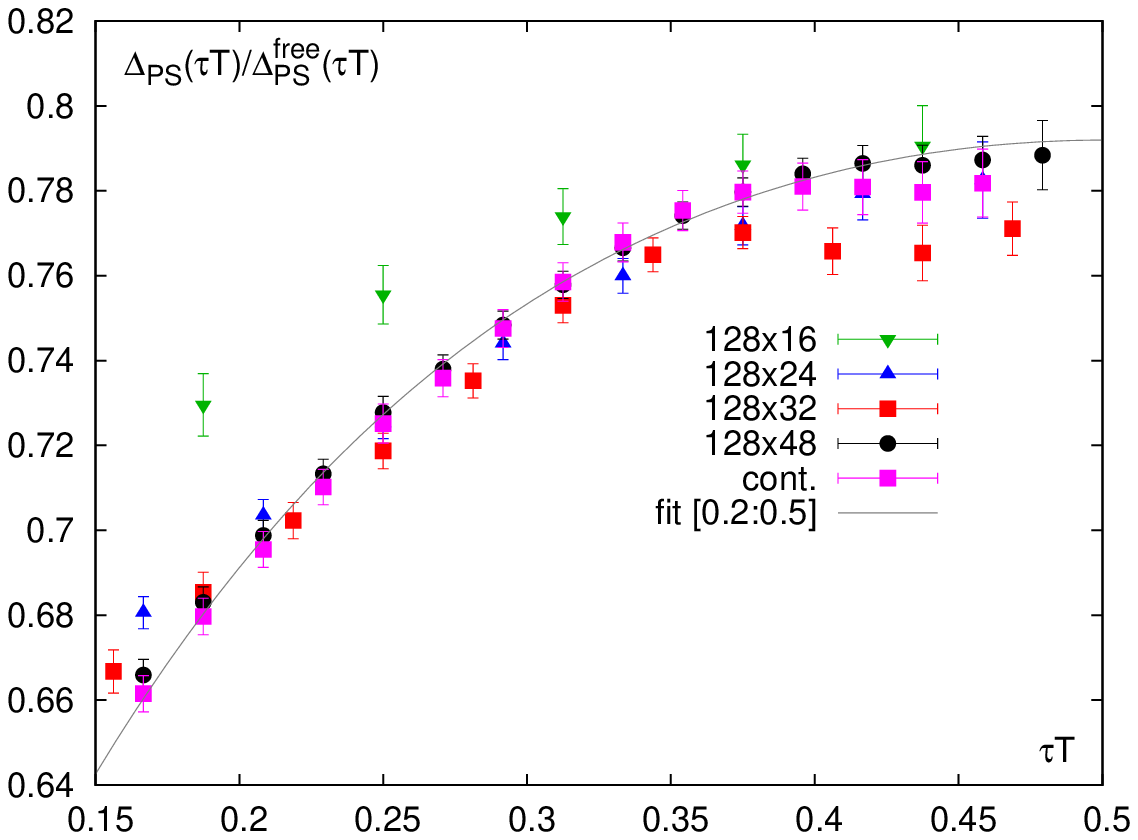}
\caption{(left): The mid-point subtracted vector correlation function normalized to the
corresponding difference for the free vector correlation function. Shown is $\Delta_V(\tau T)$ but normalized by the quark number susceptibility.
(right): The pseudo scalar case $\Delta_{PS}(\tau T)$ normalized by $G_{PS}^{(0)}$.
The Fits in both Figures obey a quartic ansatz as indicated by the definition of $\Delta_H(\tau T)$ and are shown within the interval $\tau T\in [0.2:0.5$].}
\label{curvature}
\end{figure}

In order to extract thermal moments we  examine the quantity $\Delta_H(\tau T)$
defined in Eq.~\ref{moment}. Once more we rescale this ratio of subtracted 
correlators by the 
quark number susceptibility. In Fig.~\ref{curvature} we show results obtained
from  the vector and pseudo scalar correlators, respectively.
We perform an extrapolation of $\Delta_H(\tau T)$ to the continuum,
exactly as outlined for the correlation functions themselves. 
The extrapolated data is then fitted to a quartic polynomial as indicated in 
the Taylor-expansion in Eq.~\ref{Taylor} to obtain $\Delta_H(\tau T)$ at 
$\tau T=1/2$. In the vector channel this gives
\begin{eqnarray}
\frac{G_H^{(2)}}{G_H^{(2),free}}=1.067\pm0.012\;
\label{g2g2-v}
\end{eqnarray}
where $H$ may denote either $ii$ or $V$ as noted in section 3. 
Combining these results with those of the continuum extrapolation for the
vector correlation functions 
we obtain the ratios $R_{V}^{(2,0)}$ and $R_{ii}^{(2,0)}$, 
\begin{eqnarray}
 R_{V}^{(2,0)}= 27.187\pm 0.286< R_{V,free}^{(2,0)}\;\;\;\textrm{and}\;\;\;R_{ii}^{(2,0)}= 19.217\pm0.193> R_{ii,free}^{(2,0)}\;.
\label{RVii}
\end{eqnarray}
Repeating this analysis also in the pseudo scalar channel we obtain
$\Delta_{PS}(\tau T)/G_{PS}^{(0)}$ shown in the right hand part of 
Fig.~\ref{curvature} and 
the following results for the second moment,
\begin{eqnarray}
 \frac{T^3 G_{PS}^{(2)}}{G_{PS}^{(0)}G_{PS}^{(2),free}}=0.7912\pm0.0012\;\;\;
\textrm{, thus }\;\;\; R_{PS}^{(2,0)}= 10.932\pm0.017 < R_{PS,free}^{(2,0)}
\; .
\label{g2g2-ps}
\end{eqnarray}
These results reveal some interesting properties of the individual thermal 
moments; a combination of Eqs.~\ref{g2g2-v} and  \ref{RVii}
indicates that the second thermal moment is closer to the free field value for the $H=V$ case and farther 
away for the $H=ii$, respectively.
To evaluate Eq.~\ref{g2g2-ps} in this way we need $G_{PS}^{(0)}/G_{PS}^{(0),free}$, which from Fig.~\ref{cont-corr} can be seen to be larger than 1.
Subsequently, even though we are not able to extract the latter quantity without fully controlling renormalization effects, we can conclude
the second moment must be closer to the free field limit than the zeroth.

In both channels we also tried to examine the fourth thermal moment, but our numerical results unfortunately do not permit a conclusive determination
of this value. 


\section{Electrical Conductivity}
The results obtained for the vector correlation function and its continuum 
extrapolation, as well as the result on the second thermal moment put
stringent bounds on the magnitude and shape of any contribution to the low 
frequency behavior of the vector spectral function. The small 
deviations from the free vector correlation function also suggest that 
the spectral function of the free theory is a good starting point for
an analysis of the vector spectral function at finite temperature. 
We thus used as an ansatz for the spectral function the form given
in Eq.~\ref{Ansatz}.
This ansatz depends on four temperature dependent parameters: the quark number susceptibility $\chi_q(T)$, which we already extracted from the time-like
component of the vector correlation function,
the strength ($c_{BW}(T)$) and width ($\Gamma(T)$) of the Breit-Wigner 
contribution and the higher order corrections to the high
frequency free field spectral function, which we parametrize at present by
a constant  $\kappa(T)$. Already with this ansatz we obtain good fits for 
both the spatial ($G_{ii}$) and vector ($G_V$) correlation functions. In fact, 
a combined fit to the continuum extrapolated vector correlation function in 
the Euclidean time interval $[0.25:0.5]$ and the second thermal moment, gives 
excellent results with a $\chi^2/dof$ below unity. 
For details on the fitting procedure and
a more elaborate discussion of the results we refer to \cite{karsch-francis}.
The parameters obtained using this ansatz are:
\begin{eqnarray*}
%
%
2 c_{BW}\chi_q/\Gamma=1.098\pm0.027\; ,\; \Gamma/T=2.235\pm0.075\; , \; (1+\kappa)=1.0465\pm0.003\;\;.
\end{eqnarray*}
This fit is shown in Fig.~\ref{cont-corr}(left).
Of course, as a consequence of this fit ansatz we also obtain a result 
for the behavior of the spectral function close to $\omega=0$, {\it i.e.}
we can deduce the 
electrical conductivity of the quark gluon plasma at $T \simeq 1.45T_c$:
\begin{eqnarray}
 \frac{\sigma}{T}=\frac{C_{em}}{6}\lim_{\omega\rightarrow 0}\frac{\rho_{ii}(\omega)}{\omega T}=\frac{2C_{em}}{3}\cdot\frac{c_{BW}\chi_q}{\Gamma}=(0.37\pm0.01)\cdot C_{em}\; .
\end{eqnarray}

We stress, however, that this result is a consequence of the particular ansatz
used to fit the vector correlation function. An important question is, of 
course, to what extent this ansatz is unique or allows for modifications, 
in particular at low energies, which will influence the determination of
the electrical conductivity. We intend to address this question by performing
fits within a larger class of spectral functions as well as the Maximum Entropy
Method \cite{karsch-francis}.


\section{Summary}
We have presented a detailed analysis of light meson correlation functions at $T\simeq1.45T_c$ in quenched QCD. 
For the vector current channels we find that finite volume and cut-off 
effects are under good control in a large Euclidean time interval. 
Here it is possible to take the continuum limit. 
The calculation of the second thermal moment and its inclusion in fits
greatly helped to constrain the fit parameters.
This led to an estimate of the electrical conductivity of the QGP at 
$T\simeq 1.45 T_c$, the significance of which requires further investigations
in a larger parameter space.

In the pseudo scalar channel deviations from free field behavior are 
much more pronounced. In particular the analysis of cut-off
effects is more difficult, as the 
perturbatively computed renormalization constants introduce additional 
systematic uncertainties.
However, rescaling the results by the pseudo scalar correlation
function at the midpoint yields a largely cut-off independent result. Also
finite size effects are found to be small in the pseudo scalar channel. 
This suggests that a spectral analysis of the pseudo scalar 
correlation functions should yield reliable results for its frequency
dependence and may suffer only somewhat from an imprecise knowledge of
the overall normalization. We will address the spectral analysis of
the pseudo scalar correlator elsewhere.


\acknowledgments
This work has been supported in part by contract DE-AC02-98CH10886 with the U.S. Department of Energy and by grant GRK 881 of the Deutsche Forschungsgemeinschaft.
Numerical simulations have been performed on the BlueGene/P at the New York Center 
for Computational Sciences (NYCCS) which is supported by the 
U.S. Department of Energy and by the State of New York and the BlueGene/P at the 
John von Neumann Supercomputer center (NIC) at FZ-J\"ulich,Germany.

This presentation is to a large extent based of joint work with Heng-Tong Ding,
Olaf Kaczmarek, Edwin Laermann and Wolfgang Soeldner. We thank them for all
their important input to this work.


\begin{thebibliography}{99}
\bibitem{karsch-francis} H.-T. Ding, A. Francis, O. Kaczmarek, F. Karsch, 
E. Laermann and W. Soeldner, arXiv:hep-lat/1012.4963.
\bibitem{gupta} S. Gupta, Phys. Lett. B 597, 57 (2004).
\bibitem{aarts-kim} G. Aarts, C. Allton, J. Foley, S. Hands and S. Kim, Phys. Rev. Lett. 99, 022002
(2007).
\bibitem{karsch-wyld} F. Karsch and H. W. Wyld, Phys. Rev. D 35, 2518 (1987).
\bibitem{aarts-resco} G. Aarts and J. M. Martinez Resco, Nucl. Phys. B 726, 93 (2005).
\bibitem{florkowski} W. Florkowski and B.L. Friman, Z. Phys. A347 (1994) 271.
\bibitem{wilson} K.G. Wilson, Phys. Rev. D 10 (1974), 2445.
\bibitem{lucini} B. Lucini, M. Teper and U. Wenger, JHEP 0401, 061 (2004).
\bibitem{allton} C. R. Allton, Lattice Monte Carlo data versus perturbation theory, arXiv:hep-lat/9610016.
\bibitem{luescher1} M. L\"uscher, S. Sint, R. Sommer, P. Weisz and U. Wolff, Nucl. Phys. B491
(1997) 323. 
\bibitem{luescher2} M. L\"uscher, S. Sint, R. Sommer and H. Wittig, Nucl. Phys. B491 (1997) 344.
\bibitem{sommer} M. Guagnelli and R. Sommer, Nucl. Phys. Proc. Suppl. 63, 886 (1998).
\bibitem{greeks} A. Skouroupathis and H. Panagopoulous, Phys. Rev. D 78, 119901(E) (2008);\\
 A. Skouroupathis and H. Panagopoulous, Phys. Rev. D 79, 094508 (2009)
\bibitem{karsch-wetzorke} F. Karsch, E. Laermann, P. Petreczky, S. Stickan and 
I. Wetzorke, Phys. Lett.
B 530, 147 (2002). 

\end{thebibliography}
\end{document}